\documentstyle[psfig]{mn}
\title{The Young Open Cluster NGC~2129}
\author[Carraro  at al.]
{Giovanni Carraro$^{1,2,3}$, Brian Chaboyer$^4$, and James Perencevich$^4$
\thanks{email:
gcarraro@das.uchile cl,brian.chaboyer@dartmouth.edu}\\
$^1$Departamento de Astronom\'ia, Universidad de Chile,
Casilla 36-D, Santiago, Chile\\
$^2$Astronomy Department, Yale University,
P.O. Box 208101, New Haven, CT 06520-8101 , USA\\
$^3$Dipartimento di Astronomia, Universit\`a di Padova,
Vicolo Osservatorio 2, I-35122, Padova, Italy\\
$^4$Department of Physics and Astronomy,
Dartmouth College, 6127 Wilder Laboratory
Hanover, NH 03755-3528, USA
 }

\date{\it Submitted: August 2004}
\pubyear{2005}
\begin{document}

\maketitle
\title{The open clusters NGC~2129}

\begin{abstract}

The first CCD $UBV(RI)_C$ photometric study in the area of the
doubtful open cluster NGC~2129 is presented. Photometry of a field
offset 15 arcmin northward is also provided, to probe the Galactic
disk population toward the cluster.  Using star counts, proper motions
from the UCAC~2 catalog, colour-magnitude and colour-colour diagrams we
demonstrate that NGC~2129 is a young open cluster.  The cluster radius
is 2.5 arcmin, and across this region we find evidence of significant
differential reddening, although the reddening law seems to be normal
toward its direction.  Updated estimates of the cluster fundamental
parameters are provided.  The mean reddening is found to be
E$(B-V)$=0.80$\pm$0.08 and the distance modulus is $(m-M)_o$=
11.70$\pm0.30$.  Hence, NGC~2129 is located at 2.2$\pm$0.2 kpc from
the Sun inside the Local spiral arm. The age derived from 37
photometrically selected members is estimated to be approximately 10
million years. These stars are used to provide new estimates of the
cluster absolute proper motion components.

\end{abstract}
 
\begin{keywords}
Open clusters and associations: general -- open clusters and associations:
individual: NGC~2129 -Hertzsprung-Russell (HR) diagram
\end{keywords}


\section{Introduction}

NGC~2129 (~=~OCL~467~=~C~0558+233) with $\alpha=06^{\rm h}~01^{\rm
m}.1$, $\delta=+23^{\circ} 19^{\prime}.3$ and $l=186^{\circ}.61$,
$b=+0^{\circ}.11$, J2000.0) is a clustering of stars located in the
Gemini constellation, about 2 degrees southwest of the more
conspicuous open clusters M~35 and NGC~2158 towards the Galactic
anticenter. \\ According to Trumpler (1930) the cluster is moderately
concentrated, has a wide range in star brightness, and it is
relatively rich (class {\it II 3 m }).

The group is dominated (see Fig.~1) by two close bright stars HD
250289 (LS V +23 15, V = 8.25) and HD 250290 (LS V +23 16, V = 7.36).
Within the errors, these two stars share the same proper motion
(H{\o}g et al. 2000) and radial velocity (Liu et al.\ 1989).  HD
250289 has $\mu_\alpha cos \delta =0.8\pm1.2 ~[mas/yr]$, $\mu_\delta =
-1.3\pm1.1 ~[mas/yr]$, and $V_r= 17.7\pm1.0$ [km/sec] while HD 250290
has -+$\mu_\alpha cos \delta =0.3\pm1.2 ~[mas/yr]$, $\mu_\delta =
-1.1\pm1.0 ~[mas/yr]$ and $V_r= 17.5\pm4.0$ [km/sec].  Therefore these
two stars likely constitute a pair, and are the two brightest members
of the clustering.

The first study of NGC~2129 was performed by Cuffey (1938), who
obtained photographic photometry of 111 stars and concluded that this
group of stars is a cluster with a diameter of about 5 arcmin located
570 pc from the Sun. He suggested that the two bright stars are
members of the system due to the concentration of faint stars about
them (see Fig.~1).  Hoag et al. (1961) and Johnson et al. (1961)
obtained and analyzed photoelectric/photographic UBV photometry of 45
stars down to V $\approx$ 16, concluding that the cluster is
characterized by variable extinction across its area with a mean
reddening E(B-V)=0.67 and a distance of 2.1 kpc.

Additional BV photographic photometry was secured by Voroshilov (1969)
with the same limiting magnitude, and UBV photoelectric photometry by
Lindgren \& Bern (1980, Table~V).  Both these studies, however, did
not providing information on the cluster parameters.  More recently,
Pe\~na \& Peniche (1994) performed a Stromgren ubvy-$\beta$ study of
37 stars in the region of NGC~2129 and, based upon star counts, they
concluded that there is no star clustering in the direction of
NGC~2129.  Hence, NGC~2129 might simply be a chance alignment of a few bright
stars and not a true star cluster.

To clarify this issue, we obtained deep CCD UBVRI photometry of a
field centered on NGC~2129, and CCD BVRI photometry of a control field
15 arcmin away. The size of our fields (about 8 squared arcmin) allows
us to cover the entire cluster region (see Fig.~1).

The paper is organized as follows. Section 2 describes the data
acquisition and reduction procedure. In \S 3 we compare our photometry
to previous studies.  Section 4 presents a star count analysis which
is used to determine the cluster size. In \S 5 we discuss NGC~2129
Colour Magnitude Diagrams (CMDs).  An analysis of the cluster
reddening and differential absorption is performed in \S 6. Section 7
provides estimates for the cluster age and distance. The
basic results of this investigation are highlighted in \S 8.

   \begin{figure}
   \centerline{\psfig{file=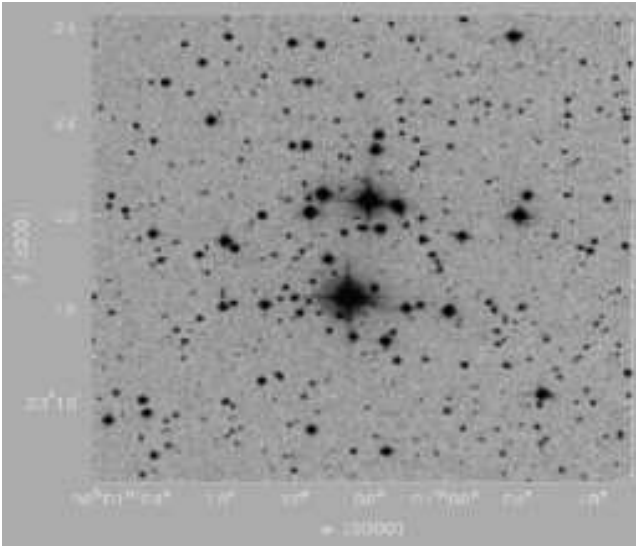,width=\columnwidth}}
   \caption{A DSS map of the observed field in the direction of NGC 2129.
The size of the field is $9 \times 9$ arcmin. North is up, East on the left.}
    \end{figure}


\section{Observations and Data Reduction}

$UBV(RI)_C$ photometry of two fields in the region of NGC~2129 were
taken at MDM observatory with the 1.3m McGraw-Hill telescope on the
nights of February 18 and 19, 2004.  The comparison field was observed
only on the first night, while the cluster field was observed 
in a similar manner both nights.  The pixel scale of the 1024 $\times$
1024 Templeton CCD is 0.50$^{\prime\prime}$, leading to a field of
view of 8.5 $\times 8.5$ arcmin in the sky.  The nights were
photometric with an average seeing of 1.4 arcsec.  We took several
short (4-12 secs), medium (40-120 secs) and long (400-1200 secs)
exposures in all the filters to avoid saturation of the brightest
stars.  Nonetheless the two bright stars HD 250289 and HD 250290 were
saturated in all exposures, and we used the photoelectric photometry
from Hoag et al.  (1961) for these two stars.

The data have been reduced with the IRAF\footnote{IRAF is distributed
by NOAO, which are operated by AURA under cooperative agreement with
the NSF.}  packages CCDRED, DAOPHOT, ALLSTAR and PHOTCAL using the
point spread function (PSF) method (Stetson 1987).  Calibration was
secured by the observation of the Landolt (1992) standard fields
PG~1047, PG~0231, SA~101, and Rubin 149 for a total of 50 standard
stars each night.  The two nights turned out to be photometrically
very similar, and therefore we decided to use all of the standard
stars in a a single photometric solution.  The calibration equations
have the following form:

\begin{eqnarray*}
u &=& U + (4.419\pm0.017) + (0.46\pm0.02) \cdot X \\
     & & {} + (0.118\pm0.026) \cdot (U-B)\\
b &=& B + (2.410\pm0.008) + (0.25\pm0.02) \cdot X \\
     & & {} + (0.037\pm0.010) \cdot (B-V)\\
v &=& V + (2.279\pm0.007) + (0.16\pm0.01) \cdot X \\
     & & {} - (0.009\pm0.009) \cdot (B-V)\\
r &=& R + (2.429\pm0.009) + (0.09\pm0.01) \cdot X \\
     & & {} + (0.061\pm0.010) \cdot (V-R)\\
i &=& I + (3.277\pm0.008) + (0.07\pm0.01) \cdot X \\
     & & {} + (0.000\pm0.009) \cdot (V-I)
\end{eqnarray*}

\noindent
and the final \textit{r.m.s.}\ of the calibration was 0.020 mag for all the 
pass-bands except for U which had an \textit{r.m.s.}\ of 0.045 mag.
The standard stars have the colour coverage  $-1.1 \leq (U-B) \leq 1.4$,
$ -0.4 \leq (B-V) \leq 1.6$, $ -0.2\leq (V-R) \leq 1.11$ 
and $ -0.3 \leq (V-I) \leq 2.3$.

   \begin{figure}
   \centerline{\psfig{file=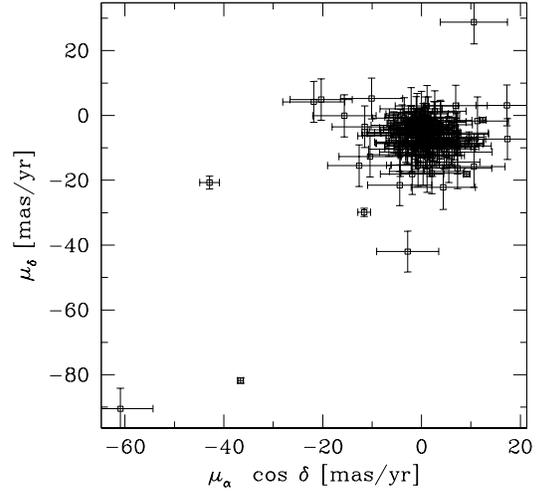,width=\columnwidth}}
   \caption{Vector Point diagram for all the stars in the field of NGC~2129 from UCAC~2}
    \end{figure}

   \begin{figure*}
   \centerline{\psfig{file=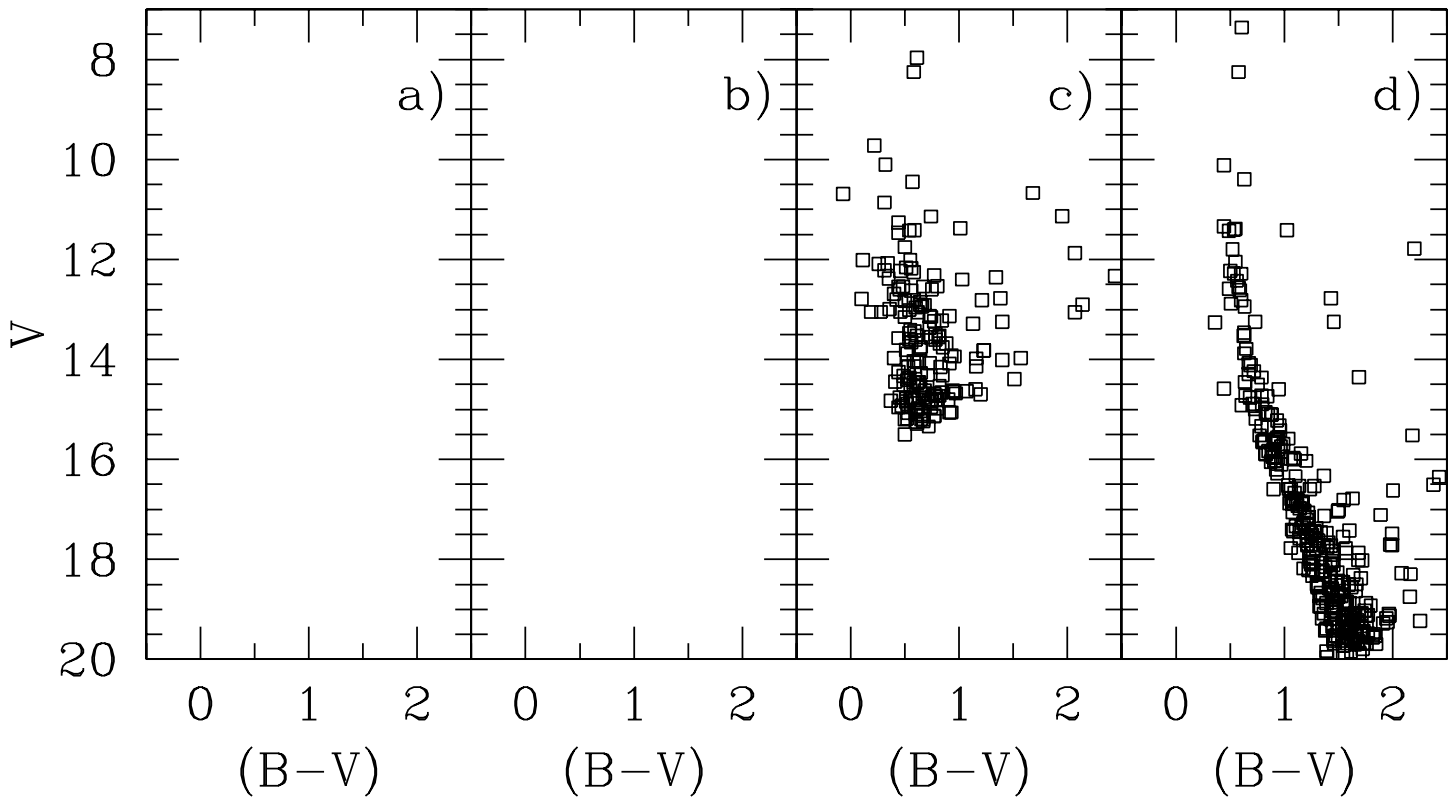}}
   \caption{
A comparison of the CMD from the present study with previous
investigation.  Panel a) presents the CMD from Hoag et al. (1961)
photoelectric photometry, panel b) the CMD from Hoag et al. (1961)
photographic photometry , panel c) the CMD from Voroshilov (1969),
and, finally, panel d) the CMD from the present study, which only 
includes stars with a colour error less than 0.07 mag.}

    \end{figure*}

Photometric errors have been estimated following Patat \& Carraro
(2001).  Stars brighter than $V \approx 20$ mag have global (internal
from DAOPHOT plus calibration) photometric errors lower than 0.15~mag
in magnitude and lower than 0.21~mag in colour.  The final photometric
data (coordinates, U, B, V, R and I magnitudes and errors) consist of
about 2,500 stars in the cluster field.

The SkyCat tool and the Guide Star Catalogue v2 (GSC-2) at ESO was
used to determine an astrometric solution for our photometry and
obtain J2000.0 coordinates for all of the stars.  There were approximately 
200 stars for which we have both the celestial
coordinates on the GSC-2 and the corresponding pixel
coordinates from our photometry. Using the IRAF tasks CCXYMATCH, CCMAP and
CCTRAN, we find the corresponding transformations between the two
coordinate systems and compute the individual celestial
coordinates for all the detected stars. The transformations have an
\textit{r.m.s.} value of $0\farcs17$, in agreement with other studies
(Momany et al. 2001, Carraro et al. 2005). 


Important information on the kinematics of the luminous stars in and
around our target can be derived from the proper motions available in
the UCAC2 catalogue (Zacharias et al. 2003).  We retrieved from the
catalogue 140 stars located in the same area of our CCD
photometry. The result is shown in the vector point diagram of
Fig.~2. There is clearly a condensation of stars centered at
$\mu_\alpha cos \delta \simeq 0~ [mas/yr]$ and $\mu_\delta \simeq -5~
[mas/yr]$, which indicates the presence of a cluster.  Recently
Beshenov \& Loktin (2003) report for 10 stars in NGC~2129 the value
$\mu_\alpha cos \delta = -0.71\pm0.28 ~[mas/yr]$ and $\mu_\delta =
-1.53\pm0.19~ [mas/yr]$ from Tycho~2.   A new estimate of
the cluster mean proper motion, based upon a new, larger 
sample of confirmed cluster stars will be presented in \S 8.
 
\section{Comparison with previous studies}

Our CCD investigation is a considerable improvement over previous
studies in the region of NGC~2129.  In Fig.~3 we compare the CMDs from
various authors with ours in the V vs (B-V) plane .  As noted before
the two bright stars HD 250289 and HD 250290 are missed in our CCD
photometry since they were saturated.  The photometry by Hoag et
al. (1961) (panel \textbf{a,b}) is clearly of much better quality than
the Voroshilov (1969) photometry (panel {\bf c)}.  The main sequence
(MS) is narrow in the Hoag et al. (1961) photometry and nicely
compares with the present study (panel {\bf d}). For this reason we
took the photometry of two bright stars from this work.  The CMD in
panel {\bf c)} does not present any distinctive feature, and from this
alone one might conclude we are viewing simply a Galactic disk field.
We note that the photometry from Cuffey (1938) is the deepest
one before the present study, although the quality of the photographic
magnitudes and colours (which are not in the standard Johnson
photometric system) is poor (see Cuffey 1938, Fig.~9).

We performed a comparison on a star by star basis with Hoag et al. (1961)
photoelectric photometry and found from 16 stars in common:

  \begin{eqnarray}
  V_{H} - V_{CC} &=& -0.04 \pm 0.05 \nonumber\\
  (B-V)_{H} - (B-V)_{CC} &=& -0.03 \pm 0.06 \nonumber\\
  (U-B)_{H} - (U-B)_{CC} &=& 0.10 \pm 0.13 \nonumber
  \end{eqnarray}

\noindent
where the suffix H stands for Hoag et al. (1961), and CC for the
present paper. A comparison to the Hoag et al. (1961) photographic
photometry (61 common stars) yields

  \begin{eqnarray}
  V_{H} - V_{CCP} &=& -0.07 \pm 0.11 \nonumber\\
  (B-V)_{H} - (B-V)_{CCP} &=& -0.04 \pm 0.09 \nonumber\\
  (U-B)_{H} - (U-B)_{CCP} &=& 0.01 \pm 0.23 \nonumber
  \end{eqnarray}

These comparisons are shown in Fig.~4 and 5.  The photoelectric
photometry agrees well with our photometry to V = 12.5, but fainter
than this magnitude some scatter starts to be present (Fig.~4).  There
appears to be systematic offsets between the datasets in Fig.~5,
although the sizes of the differences are reasonable in a comparison
between CCD and photographic photometry.  Finally, we notice that
there appears to be a colour term effect in the U band, which runs in
opposite direction in the photoelectric (Fig~4) and photographic
(Fig.~5) photometry.

   \begin{figure}
   \centerline{\psfig{file=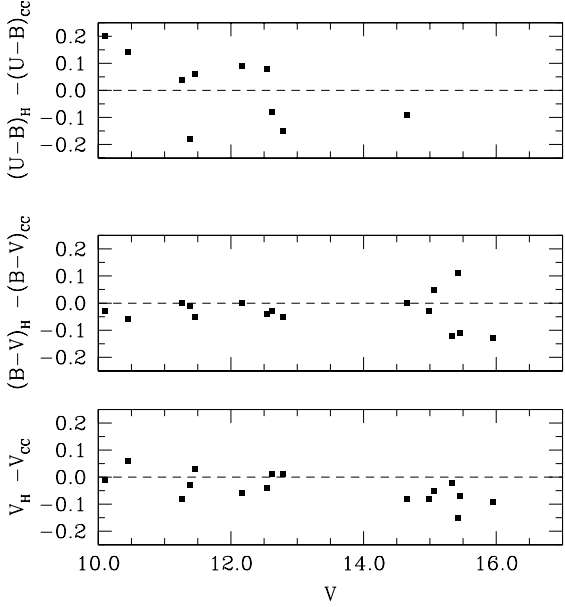,width=\columnwidth}}
   \caption{Comparison between our photometry and the photoelectric photometric
by Hoag et al. (1961).}
    \end{figure}

   \begin{figure}
   \centerline{\psfig{file=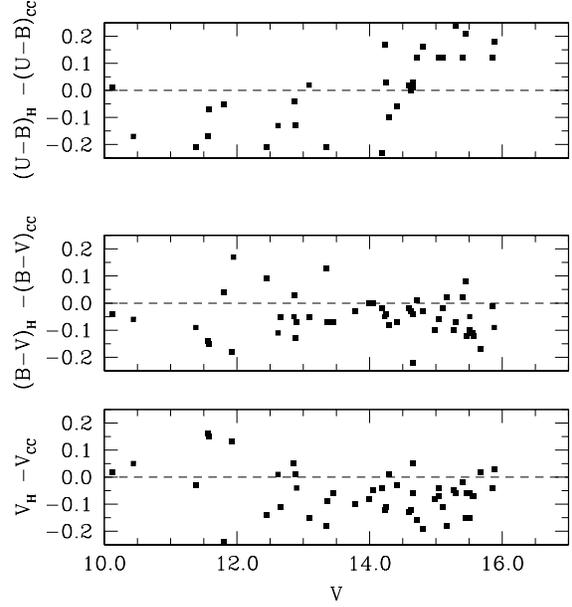,width=\columnwidth}}
   \caption{Comparison between our photometry and the photographic photometric
by Hoag et al. (1961).}
    \end{figure}

\section{Star counts and cluster size}

The cluster radius is one of the most important cluster parameters,
useful, together with cluster mass, for a determination of cluster
dynamical parameters.  The aim of this section is to obtain the
surface density distribution of NGC 2129, and derive the cluster size
by means of star counts. Star counts allow us to determine statistical
properties of clusters with respect to the surrounding stellar
background.  

   \begin{figure}
   \centerline{\psfig{file=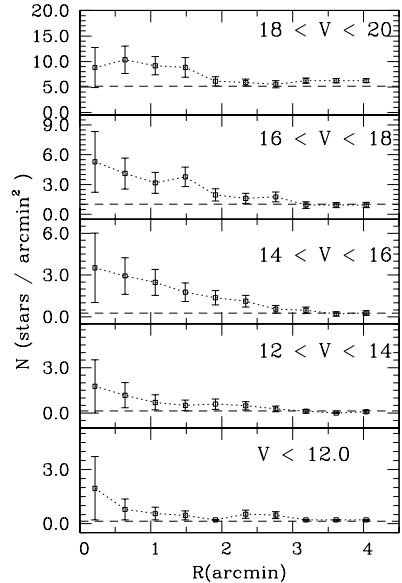,width=\columnwidth}}
   \caption{Star counts as a function of radius from the adopted cluster
center for various magnitude intervals. The dashed line in each panel
indicates the mean density level of the surrounding Galactic disk
field at that magnitude level.}
    \end{figure}

   \begin{figure*}
   \centerline{\psfig{file=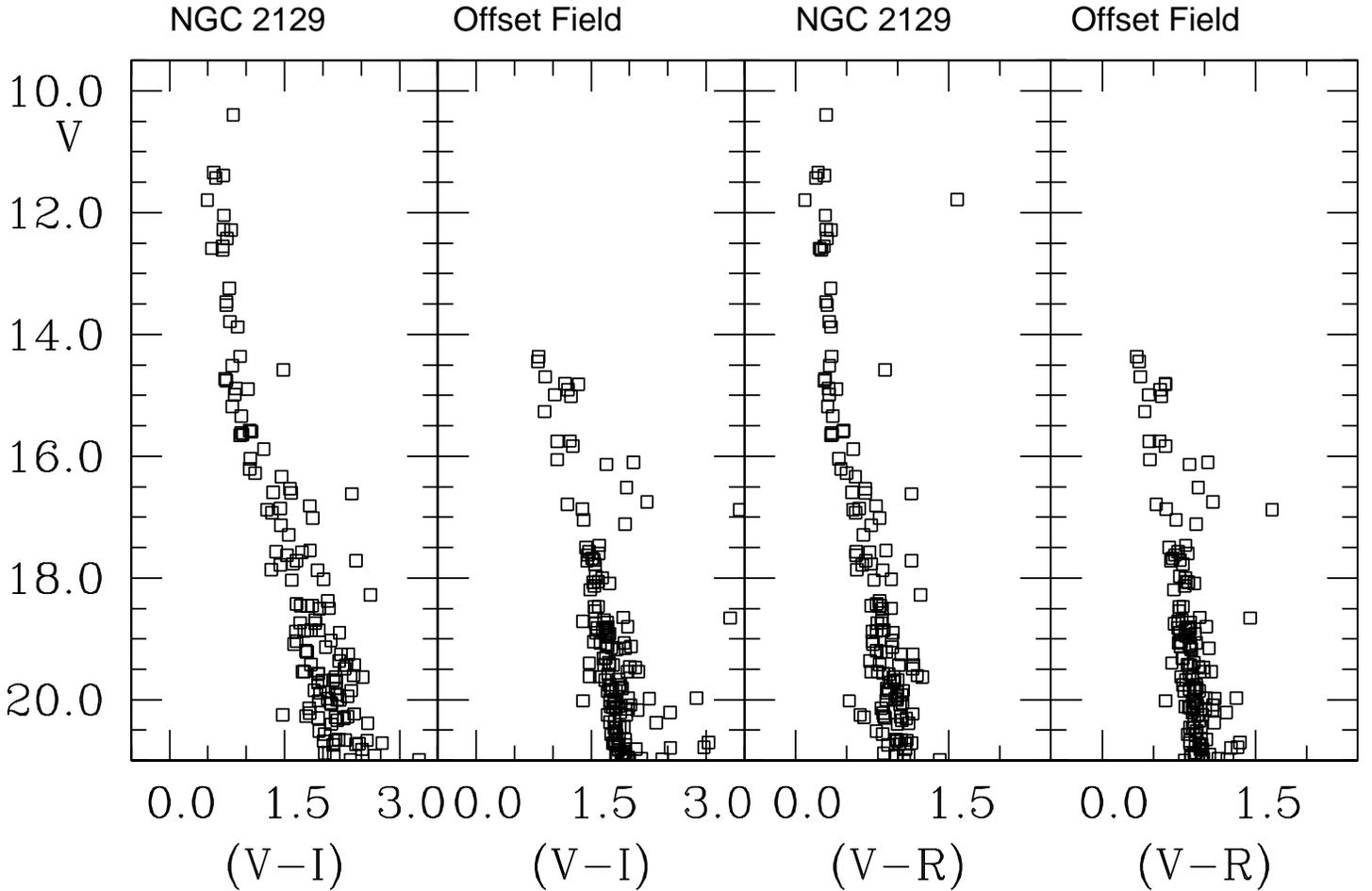}}
   \caption{CMDs for the stars in the field of NGC~2129 and in the off-set field.}
    \end{figure*}

In Fig.~1, NGC 2129 appears as a concentration of bright stars in a
region of about 4 arcmin diameter.  To derive the radial
stellar surface density we determine the highest peak in the stellar
density to find the cluster center.  The two bright stars HD 250289
and HD 250290 have been included in this, and the following star count
calculations.  The adopted center is placed at $\alpha = 06:01:07.0$;
$\delta = +23:19:18.0$, similar to that given by Dias et al. (2002).

The radial density profile is constructed by performing star counts
inside increasing concentric annuli $0\farcm5$ wide, around the
cluster center and then dividing by their respective surface
areas. This is done as a function of apparent magnitude, and compared
with the mean density of the surrounding Galactic field in the same
brightness interval. The contribution of the field has been estimated
through star counts in the region outside 6 arcmin from the cluster
center. Poisson standard deviations have been computed and normalized
to the area of each ring as a function of the magnitude, both for the
cluster and for the field.  The result is shown in Fig.~6, where one
readily sees that NGC~2129 emerges significantly from the mean field
for magnitudes brighter than V$\approx$18.  At fainter magnitudes the
cluster starts to mix with the Galactic disk population.  Based on the
radial density profiles in Fig.~6, we find that stars brighter than
V=20 provide a cluster radius of approximately 2 arcmin.  We adopt as
a final estimate of the radius $2.0\pm0.5$ arcmin. This value of the
cluster radius is adopted throughout this work and is in basic
agreement with the estimate of 2.5 arcmin reported by Cuffey (1938),
which was simply based on visual inspection.  We stress however that
this radius is not the limiting radius of the cluster, but the
distance form the cluster center at which the cluster population
starts to be confused with the field population. At odds with Pe\~na
\& Peniche (1994), we find that NGC~2129 appears as a clear star
cluster standing above the mean Galactic field.

\section{Analysis of the CMDs}

Further confirmation of the cluster nature of NGC~2129 can be found in
the comparison of the CMD for the cluster and the control field region
depicted in Fig.~7.  In this figure we show the V vs (V-I) and V vs
(V-R) CMDs of NGC~2129 star within 2.5 arcmin from the cluster center,
and the same CMDs for stars in a similar area region taken from the
control field.  We considered only the stars brighter than V = 21 and
with $\sigma_V \leq 0.05$ .  It is readily seen that the cluster
actually exists, and it exhibits a nice MS extending from V = 10 down
to V = 21.  On the other hand, the field MS sharply stops at V
$\approx$ 17.5, and contains only an handful of brighter stars.  The
cluster MS presents some scatter, larger than expected from the sole
photometric errors (see \S 2), and which we ascribe mostly to
differential reddening across the cluster area (see next Section).  This
is not unexpected, due to the cluster position, low in the Galactic
thin disk.

\section{Individual reddenings, membership and differential reddening}
We use $UBV$ photometry to derive stars' individual reddenings and
membership.   Briefly, individual reddening values have been computed by
means of the usual reddening free parameter $Q$:

\[
Q = (U-B) - 0.72  \times (B-V) - 0.05 \times (B-V)^2              ,
\]

\noindent
and the distribution of the stars in the two colour diagram.  We
follow the procedure described in Carraro (2002), Ortolani et
al. (2002) and Baume et al. (2005), where the young open clusters
Trumpler~15, NGC~1220 and Markarian~50 were studied.  This method is a
powerful one to isolate early spectral type (from $O$ to $A5$) stars
having common reddening, which are most probably cluster members (see
also, for a reference, the study of Trumpler~14 by Vazquez et
al. (1996).  Note that the reddening based membership selection gives
similar results to proper motion based membership selection (see
Cudworth et al. (1993) and Patat \& Carraro (2001) for some
applications to star clusters in the Carina region).

Our results are shown in Fig~8, where we plot all the stars brighter
than V = 19 having $UBV$ photometry in the two-colour Diagram.  In
this plot the solid line is an empirical ZAMS taken from Schmidt-Kaler
(1982).  The bulk of the stars are confined within a region defined by
two ZAMS shifted by $E(B-V)$~=~0.60 and 0.90 (dashed lines),
respectively.  The 37 stars (filled symbols) which occupy this region
have a mean reddening $E(B-V)~=~0.80\pm0.08 (r.m.s.)$.  Since the
spread in reddening is larger than the photometric errors, which are
typically $\Delta(U-B) \approx \Delta(B-V) \approx 0.03$, this
indicates the presence of differential reddening across the cluster,
as suggested also by Cuffey (1938).  This technique has been applied
to all the stars but for HD 250289 and HD 250290, which are stars with
spectral type B2IIIe and B3I (Morgan et al. (1955) and Hoag \&
Applequist (1965)) and are probably evolved stars.  For these two
stars absolute magnitudes and colours have been derived from Wegner
(1994), assuming their spectral types and luminosity classes.

For some stars, an unambiguous reddening solution is not possible, and
these stars are plotted with open symbols.  A reddening solution is
not possible since these stars are located in a region where larger
reddening ZAMS cross the E$(B-V)$ =0.90 mag ZAMS, making it impossible
to effectively disentangle members from non members.

   \begin{figure}
   \centerline{\psfig{file=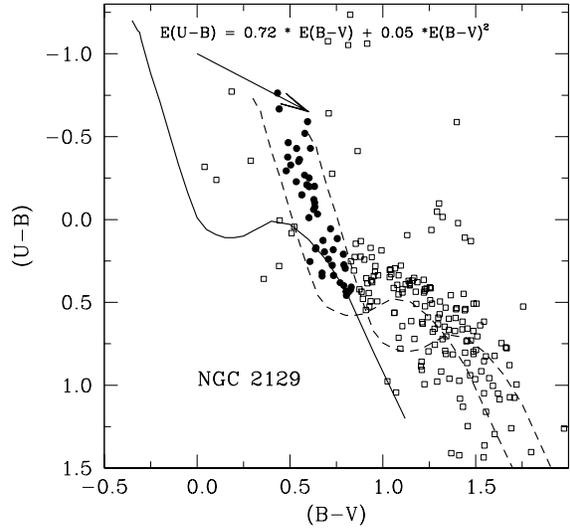,width=\columnwidth}}
   \caption{
UBV Colour-colour diagram for all the stars in the field of NGC~2129
having $UBV$ photometry.  The solid line is the Schmidt-Kaler (1982)
empirical ZAMS, whereas the dashed lines are the same ZAMS, but
shifted by E$(B-V)$~=~0.60 and E$(B-V)$~=~0.90, respectively. Filled
symbols indicate stars having reddening in the range $0.60 \leq E(B-V)
\leq 0.90$, for which an unambiguous reddening solution has been
possible.}
    \end{figure}

    \begin{figure}
   \centerline{\psfig{file=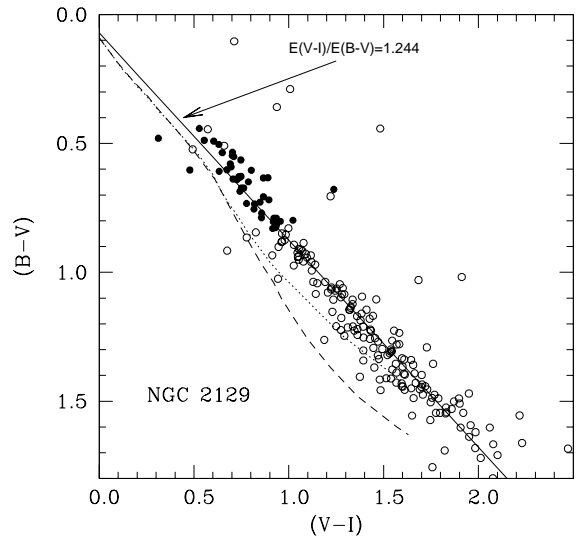,width=\columnwidth}}
   \caption{
BVI Colour-colour diagram for all the stars in the field of NGC~2129
having $UBV$ photometry.  The solid line is the normal reddening
vector.  whereas the dotted and dashed lines are the intrinsic
positions for stars of luminosity classes V and II, respectively,
taken form Cousins 1978a,b.  Only stars for which an unambiguous
reddening solution has been possible are plotted. Symbols as in
Fig.~8.}

    \end{figure}

In summary, there are 37 photometric members, for which we
can derive estimates of the distance and age of the cluster.  The member 
stars are listed in Table~1, together with coordinates, magnitudes, colours
and proper motion component from UCAC2.  We derived
the photometric spectral type from the $Q$ parameter (columns 10 and
11 in Table~1).  These spectral types compare very well with what we
can found in the literature.  In fact, Hoag \& Applequist (1965)
report for stars 3, 4 and 7 spectral type of B3 III, B6 and B5, very
close to our determinations.  In addition, McCuskey (1967) found that the 7
brightest stars in the field of NGC~2129 are of OB spectral type and
Chargeishvili (1988) reports for a few bright stars spectral
classification from an objective prism analysis, suggesting that they
are of B spectral type.

In Fig.~9 we plot all the stars having UBV photometry in the (V-I) vs
(B-V) diagram. In this plot the solid line is the normal reddening
vector from Dean et al. (1978), whereas the dotted and dashed lines
are the intrinsic positions for stars of luminosity classes V and II,
respectively, taken form Cousins 1978a,b .  Given that the stars
follow the standard reddening vector, the ratio of total to selective
absorption [$R_V = A_V/E(B-V)$] is basically normal.

\begin{table*}
\tabcolsep 0.11cm
\caption{Basic data  of photometric likely
member stars in the field of NGC~2129.}
\begin{tabular}{cccccccccccccc}
\hline
\hline
\multicolumn{1}{c}{ID} &
\multicolumn{1}{c}{Hoag et al.(1961)} &
\multicolumn{1}{c}{RA} &
\multicolumn{1}{c}{DEC} &
\multicolumn{1}{c}{V} &
\multicolumn{1}{c}{$(B-V)$} &
\multicolumn{1}{c}{$(U-B)$}  &
\multicolumn{1}{c}{$(V-R)$}  &
\multicolumn{1}{c}{$(V-I)$}  &
\multicolumn{1}{c}{E$(B-V)$} &
\multicolumn{1}{c}{$Q$} &
\multicolumn{1}{c}{{\it Sp.Type}} &
\multicolumn{1}{c}{$\mu_{\alpha}$}&
\multicolumn{1}{c}{$\mu_{\delta}$} \\
\hline
   4& 3  &6:01:05.25&   +23:21:24.85&   10.115&    0.442&   -0.516&    0.245&    0.528&  0.707&    -0.695& B2  & 0.0$\pm$0.9&    -2.3$\pm$0.7 \\
   5& 4  &6:01:17.50&   +23:19:30.80&   11.391&    0.547&   -0.351&    0.283&    0.721&  0.812&    -0.660& B3  &-0.8$\pm$1.0&    -1.7$\pm$0.9 \\
   6& 7  &6:00:59.64&   +23:17:57.11&   11.431&    0.490&   -0.463&    0.198&    0.603&  0.779&    -0.629& B3  & 0.2$\pm$1.0&    -7.6$\pm$0.6 \\
   7&    &6:01:04.81&   +23:17:20.00&   11.400&    0.535&   -0.429&    0.200&    0.649&  0.824&    -0.639& B3  & 0.8$\pm$6.5&    -0.6$\pm$6.3 \\
   9&    &6:00:52.14&   +23:16:05.46&   11.794&    0.523&    0.042&    0.090&    0.493&  0.649&    -0.347& B7  & 4.4$\pm$6.5&    22.2$\pm$6.8\\
  10&    &6:01:03.04&   +23:18:02.17&   12.043&    0.551&   -0.362&    0.291&    0.710&  0.821&    -0.674& B3  &-1.7$\pm$3.7&    -2.1$\pm$1.4\\
  12&    &6:01:14.43&   +23:18:08.68&   12.277&    0.534&   -0.227&    0.299&    0.703&  0.755&    -0.627& B3  & 3.9$\pm$6.4&    -2.9$\pm$7.2\\
  13&    &6:01:03.46&   +23:20:09.15&   12.286&    0.603&   -0.198&    0.345&    0.801&  0.832&    -0.652& B3  & 0.6$\pm$1.0&    -1.0$\pm$0.8\\
  14& 8  &6:00:58.45&   +23:19:32.38&   12.224&    0.504&   -0.329&    0.253&    0.631&  0.751&    -0.705& B2  &-0.7$\pm$1.4&     1.2$\pm$0.8\\
  15&    &6:01:04.94&   +23:21:43.33&   12.426&    0.563&   -0.149&    0.306&    0.747&  0.765&    -0.571& B4  &-2.5$\pm$2.6&    -2.4$\pm$1.3\\
  16& 10 &6:01:10.84&   +23:21:31.74&   12.586&    0.488&   -0.376&    0.233&    0.553&  0.747&    -0.640& B3  & 1.1$\pm$0.6&    -2.5$\pm$0.6\\
  17& 11 &6:01:10.88&   +23:15:28.03&   12.614&    0.591&   -0.210&    0.254&    0.695&  0.820&    -0.653& B2  & 0.8$\pm$6.3&    -9.5$\pm$6.3\\
  18&    &6:01:17.80&   +23:18:07.07&   12.552&    0.578&   -0.267&    0.278&    0.690&  0.824&    -0.701& B2  & 2.7$\pm$6.3&     1.3$\pm$6.3\\
  19&    &6:01:11.77&   +23:23:42.82&   12.884&    0.509&    0.082&    0.262&    0.659&  0.619&    -0.298& B8  &-1.3$\pm$1.9&    -1.0$\pm$1.9\\
  20&    &6:01:10.63&   +23:20:05.94&   12.941&    0.634&   -0.075&    0.339&    0.866&  0.829&    -0.552& B4  &-0.6$\pm$1.8&    -2.9$\pm$1.8\\
  21&    &6:01:16.75&   +23:19:20.91&   12.810&    0.601&   -0.010&    0.253&    0.673&  0.766&    -0.462& B5  &-0.6$\pm$1.8&    -5.5$\pm$1.3\\
  24&    &6:01:01.59&   +23:19:29.59&   13.515&    0.632&   -0.199&    0.308&    0.740&  0.868&    -0.675& B2 &-10.1$\pm$6.3&     5.2$\pm$6.3\\
  25&    &6:01:18.46&   +23:22:04.92&   13.242&    0.732&    0.182&    0.344&    0.777&  0.867&    -0.372& B7  &-1.7$\pm$1.9&    -1.6$\pm$1.9\\
  26&    &6:01:08.42&   +23:17:53.81&   13.461&    0.628&   -0.121&    0.296&    0.737&  0.838&    -0.594& B3  & 0.2$\pm$2.0&    -4.6$\pm$2.0\\
  28&    &6:01:09.50&   +23:23:39.74&   13.876&    0.633&   -0.102&    0.345&    0.890&  0.837&    -0.578& B4  &-1.3$\pm$1.9&    -1.0$\pm$1.9\\
  30&    &6:01:08.02&   +23:18:19.20&   13.529&    0.627&   -0.059&    0.289&    0.748&  0.816&    -0.532& B4  & 0.2$\pm$2.0&    -4.6$\pm$2.0\\
  31&    &6:00:53.55&   +23:19:58.22&   13.787&    0.648&   -0.032&    0.327&    0.787&  0.833&    -0.521& B4  & 0.6$\pm$0.6&    -1.3$\pm$0.6\\
  34&    &6:01:05.91&   +23:20:13.47&   14.068&    0.672&    0.341&    0.196&    0.762&  0.738&    -0.166& B9  &-1.1$\pm$1.9&    -5.9$\pm$1.9\\
  35&    &6:01:14.49&   +23:20:03.71&   14.234&    0.718&    0.056&    0.334&    0.894&  0.892&    -0.487& B5 &-12.6$\pm$6.4&   -15.5$\pm$6.4\\
  38&    &6:01:08.12&   +23:18:13.08&   14.290&    0.672&    0.321&    0.250&    0.750&  0.744&    -0.185& B9  & 2.7$\pm$6.3&     1.3$\pm$6.3\\
  42&    &6:01:11.95&   +23:21:31.40&   14.449&    0.638&    0.178&    0.323&    0.706&  0.749&    -0.302& B8  & 1.1$\pm$0.6&    -2.5$\pm$0.6\\
  49& 17 &6:00:48.13&   +23:19:11.53&   14.760&    0.686&    0.194&    0.282&    0.741&  0.804&    -0.323& B8  &-1.1$\pm$6.3&   -10.2$\pm$6.4\\
  50&    &6:01:07.28&   +23:18:13.54&   14.886&    0.707&    0.238&    0.324&    0.865&  0.816&    -0.296& B8  & 2.7$\pm$6.3&     1.3$\pm$6.3\\
  55&    &6:01:07.89&   +23:19:39.38&   14.919&    0.607&    0.253&    0.294&    0.632&  0.685&    -0.203& B9  &-4.4$\pm$6.5&   -21.5$\pm$6.3\\
  56&    &6:01:06.40&   +23:22:34.49&   14.727&    0.640&    0.170&    0.287&    0.726&  0.754&    -0.312& B8  & 7.2$\pm$6.3&    -5.4$\pm$6.3\\
  57&    &6:00:57.93&   +23:24:09.39&   14.987&    0.728&    0.276&    0.329&    0.847&  0.830&    -0.275& B8  & 1.7$\pm$6.4&   -12.4$\pm$6.4\\
  62&    &6:00:53.78&   +23:18:00.92&   15.179&    0.734&    0.336&    0.315&    0.819&  0.818&    -0.219& B9  &-1.7$\pm$3.7&    -2.1$\pm$1.4\\
  68& 22 &6:00:54.18&   +23:19:58.06&   15.336&    0.790&    0.399&    0.366&    0.933&  0.867&    -0.201& B9  & 0.6$\pm$0.6&    -1.3$\pm$0.6\\
  72& 24 &6:01:05.90&   +23:20:30.71&   15.518&    0.770&    0.381&    0.345&    0.857&  0.848&    -0.203& B9  &-1.1$\pm$1.9&    -5.9$\pm$1.9\\
  81&    &6:01:00.26&   +23:18:50.96&   15.643&    0.801&    0.434&    0.360&    0.954&  0.870&    -0.175& A0.5 &-0.2$\pm$6.4&    -4.2$\pm$6.3\\
  84&    &6:01:12.76&   +23:18:30.40&   15.619&    0.816&    0.437&    0.354&    0.934&  0.887&    -0.184& A0.5 & 3.9$\pm$6.4&    -2.9$\pm$7.2\\
\hline
\end{tabular}
\end{table*}

  \begin{figure}
   \centerline{\psfig{file=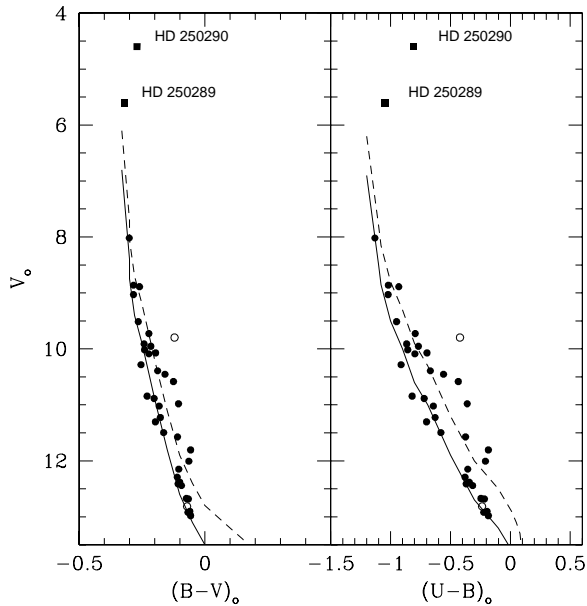,width=\columnwidth}}
   \caption{
Reddening corrected CMDs for NGC~2129. Superposed is an empirical ZAMS
taken from Schmidt-Kaler (1982, solid line) shifted by
(m-M)$_{o,V}$=11.70. The dashed line is the same ZAMS 0.70 mag
brighter, which mimics the position of unresolved binary stars.}
    \end{figure}

\section{age and distance}

The reddening corrected CMDs for the likely member stars from Table~1
are plotted in Fig.~1..  We have superposed the empirical
Schmidt-Kaler (1982) ZAMS (solid line), shifted by $(m-M)_o
=11.7\pm0.2$ mag, which provides a nice fit to the observed
distribution of stars.  A 0.70 mag brighter ZAMS (dashed line) is
shown to mimic the location of unresolved binary stars.  The two
brightest stars HD 250289 and HD 250290 are plotted as solid
squares. The open circle is a star ($\#9$) that was found to be a
photometric member, but whose proper motion components differ
significantly from the mean.  Therefore we consider it as a foreground
star. Its position in the CMDs supports this conclusion.

The absolute distance modulus implies that NGC~2129 is located
$2200\pm200$ pc from the Sun.  The relatively large uncertainty is due
to the difficulty of fitting the almost vertical structure of the
MS. The Galactocentric coordinates are $X = -250\,$pc, $Y = 2200\,$pc,
$Z = 4\,$ pc and the Galactocentric distance is $\mathrm{R_{GC}} =
10.70\,$kpc.  The cluster lies in the extension of the Local spiral
arm towards the Galactic Third Quadrant.  The distance modulus we find
is in nice agreement with Johnson et al. (1961), who found
$(m-M)_0=11.60$ and a distance of 2100 pc.

The two brightest stars HD 250289 and HD 250290 are of spectral type
B2III and B3I and are likely evolved stars.  On the other hand, the
star at $V \approx 8.1$ appears to be on the main sequence with an
absolute magnitude of $\mathrm{M_V}= -3.6$, and an estimated spectral
type of approximately B2-B3.  The two colour diagram (Fig.~8)
enables one to estimate the spectral classification of stars 
with spectral types ranging from B2 to A5 (see Table 1).  
If the stars having $B2-B3$ spectral type are
still along the MS, we infer that the age of NGC~2129 is around 10 Myrs.

\section{Conclusions}

In this paper we have presented the first CCD multicolour photometry
for the stars in the field of NGC~2129, and provide the first estimate
of its fundamental parameters.  Our findings can be summarized as
follows:

\begin{itemize}

\item  NGC~2129 is a compact group of stars with a radius of 2.0-2.5
arcmin, or 1.0-1.3 pc at the distance of the cluster;

\item we identified 37 likely members with spectral type earlier than
$A5$ on the basis of reddening, proper motion and the position in the reddening
corrected CMDs;

\item the cluster is situated about 2200 pc away from the Sun in the
anticenter direction, inside the Local spiral arm;

\item the mean reddening is $E(B-V)=0.82\pm0.08$, and there is 
substantial differential reddening;

\item the probable age of NGC~2129 is approximately 10 Myrs;

\item the mean proper motion components of the 37 cluster members are  
          $\mu_\alpha cos \delta = 1.03\pm0.52~[mas/yr]$  and 
          $\mu_\delta = -3.32\pm0.90~[mas/yr]$  .
    
\end{itemize}

\section*{Acknowledgements}
GC is profoundly indebted to Brian Skiff for 
providing numberless very useful comments and suggestions.
The work of GC is supported by {\it Fundaci\'on Andes}.  Research
supported in part by a NSF CAREER grant 0094231 to BC.  BC is a
Cottrell Scholar of the Research Corporation.

\end{document}